\documentclass[english]{revtex4-1}
\usepackage[T1]{fontenc}
\usepackage[latin9]{inputenc}
\usepackage{color}
\usepackage{float}

\makeatletter

\providecommand{\tabularnewline}{\\}

\usepackage{breakcites}

\makeatother

\usepackage{babel}
\begin{document}

\title{Supersymmetric Extension of Preonic Models: General Remarks}

\author{B. B. Oner}

\address{Department of Material Science and Nanotechnology, TOBB University
of Economics and Technology, Ankara, Turkey}
\email{b.oner@etu.edu.tr}

\author{S. Sultansoy}

\address{TOBB University of Economics and Technology, Ankara, Turkey}

\address{ANAS Institute of Physics, Baku, Azerbaijan}
\email{ssultansoy@etu.edu.tr}

\begin{abstract}
We present some general remarks on supersymmetric extensions of fermion-scalar
and three-fermion preonic models with an assumption of supersymmetry
is realized at preonic level. The motivation and the requirement of
this assumption are briefly given. In general, supersymmetric extensions
of fermion-scalar (three-fermion) models predict 4 (10) ``super''
partners for each Standard Model fermion.
\end{abstract}
\maketitle

\section{Introduct\i on}

Historical development of the search of the ``fundamental'' matter(s)
and interaction(s) shows us that pre-judgements on the fundamentality
of the matter usually turned out to be wrong. The time has shown that
most of the previously tought fundamental matter could be accepted
elementary only on a lower energy region. Electron, which was already
known due to cathode ray experiments in 1897 \cite{key-Thomson},
evoked the idea of compositeness of the atom. Ernest Rutherford revealed
its real structure \cite{key-Rutherford} and suggested to increase
collision energies for a better understanding of the fundamentality.
A few years later, in 1919, proton is discovered \cite{key-Proton}
and, therefore, the nuclei are also proved to be composites. Discovery
of zero electric charge nutron has been a little late. In 1932, neutron
was tought to be a composite particle that consists of proton and
electron \cite{key-Chadwick-neutron}. Even though neutrons are composites,
the real story was again different. After 1950s, hadronic structure
of nuclei was revealed and in the next decade quarks were introduced
as building blocks of hadrons \cite{key-Gell-MannZwei,key-Gell-MannZwei2,key-Gell-MannZwei3}.
Baryons, like proton and neutron, are proved to consist of three quarks
while mesons, like pi-meson, are quark-antiquark composites. After
these discoveries with other enormous number of experiments, finally,
Standard Model (SM) is developed and took its final form in 1970s
\cite{key-SM}. SM is experimentally confirmed in low energy region
totally after the discovery of the 125 GeV Higgs boson in 2012 \cite{key-Higgs}.

Most of previously \textit{fundamental} questions in physics found
their answers within the brief history given above. However, the whole
story did not end since SM still has critical deficincies to explain
some important phenomena. We present a detailed list of these questions/problems
below \cite{key-fourRemarks}:

Why does the most-fundamental model need large number of free observable
parameters (26 if neutrinos are Dirac particles, 30+? for Majorana
neutrinos) put by hand?

What is the reason of the number of the SM generations?

What is the reason for the mass hierarchy between these generations?

What is the reason of the quark-lepton symmetry? Are the neutrinos
Dirac or Majorana particles?

Why do the CKM angles and weak mixing angles take their values?

Why do the electric charge quantizations of quarks and leptons are
related ($Q_{u}=+2Q_{e}/3,\;Q_{d}=-Q_{e}/3$) ?

Why is L-R symmetry broken in the SM and why do quarks and leptons
share similarity in the weak sector?

What is the reason of matter-antimatter asymmetry?

What is the real origin of CP-violation?

Why are all the known interactions built on a gauge symmetry? 

What is origin of ``confinement'' of colored objects? Are they ``truly
confined''?

How can be the role of gravity put in its place in unification theories?

How can one solve theoretical problem of quadratic divergence in the
Higgs mass?

...

Different phenomenological and theoretical approaches beyond the Standard
Model (BSM) were proposed in order to overcome at least a part of
the problems encountered. Composite models, super-symmetry (SUSY),
and\textcolor{red}{{} }grand unified models and extra dimensions can
be regarded as the most efficient ones among these proposals. However,
no experimental evidence has been found supporting these ideas yet.
Therefore, we know that if a new physics exists, the new physics scale,
$\Lambda$ , is certainly over a few TeV. We present brief histories
and necessities of composite models and SUSY below.

\textit{Composite Models}: Leptons and quarks in SM are assumed to
be elementary particles which seems questionable especially considering
the mass and mixing patterns of these SM fermions. Quantization of
their electric charges and their similar chiral structures can be
added to the list. In this manner, composite models seem to be the
strongest candidates since they are able to clarify these questions
naturally. Historical arguments favoring composite models are presented
in Table I.
\begin{center}
\begin{table}[H]
\caption{Examples of \textquotedblleft fundamental\textquotedblright{} particle
inflations encountered in the last century. }

\centering{}%
\begin{tabular}{|c|c|c|c|}
\hline 
Stages & 1870s-1930s & 1950s-1970s & 1970s-2020s\tabularnewline
\hline 
\hline 
Fundamental Particle Inflation & Chemical Elements & Hadrons & Quarks \& Leptons\tabularnewline
\hline 
Systematic & Periodic Table & Eight-fold Way & Family Replication\tabularnewline
\hline 
Confirmed Predictions & New Elements & New Hadrons & $l_{8}$ and $q_{6}$?\tabularnewline
\hline 
Clarifying Experiments & Rutherford & SLAC DIS & LHC? or rather FCC?\tabularnewline
\hline 
Building Blocks & Proton, Neutron, Electron & Quarks & Preons?\tabularnewline
\hline 
Energy Scale & MeV & GeV & Multi-TeV?\tabularnewline
\hline 
Impact on Technology & Exceptional & Indirect & Exceptional?\tabularnewline
\hline 
\end{tabular}
\end{table}
\par\end{center}

Composite models have been developed since forty years ago (see Ref.
\cite{key-preonsBook} and references therein). Pati and Salam denoted
the fundamental particles as PRE (-entities) first \cite{key-preEntity}
and later this name evolved as preons \cite{key-patiSalamPreon}.
For detailed reviews of composite models see Refs. \cite{key-Terazawa1,key-Terazawa2,key-Terazawa3}.
Even though most of the models present consistent arguments to the
abovementioned SM problems, they did not arouse interest as much as
they deserved; essentially compared to the SUSY. 

\textit{SUSY}: One of the most promising sides of the SUSY is cancellation
of quadratic divergent in the Higgs mass, $m_{H}$, by opposite-sign
contributions of superpartners. For a detailed review of SUSY and
some other important supporting ideas, see \cite{key-haberKane,key-SUSYguncel}.
However a critical problem arises: number of observable free parameters.
Even in minimal super-symmetric Standard Model (MSSM), this number
increases up to more than 200 \cite{key-fourRemarks,key-mssm=0000231,key-mssm=0000232,key-mssm=0000233}
while it is only 26 in SM itself. Another critical issue is that supersymmetric
partners of SM fermions have been expected to have masses below 1
TeV \cite{key-haberKane} in order to preserve $m_{H}=125$ GeV, however,
most of this region is already excluded by LHC \cite{key-LHCsusyExc,key-LHCsusyExc2}.

To sum up, compositeness and SUSY, seperately, have lack of ability
to explain most of the SM problems and these problems seem to remain
unsolved unless proposed BSM models are combined. One should note
that the SM is rather a theory of fundamental interactions than fundamental
fermions \cite{key-23}. Therefore, it is natural to consider an option
in which SM fermions are composites and SUSY is realized at a more
fundamental level. 

Pati and Salam proposed supersymmetry at preonic or pre-preonic level
in 1982 \cite{key-PatiSalamSUSYPreonModel} with an assumption of
the lepton number as the fourth color \cite{key-patiSalam1974leptonfourthcolor}.
Only limited number of studies handled preonic SUSY after 1980s and\textcolor{red}{{}
}there is not any fully-realistic preonic model yet. This is especially
because of the unknown preonic level interactions. However, one can
make some general phenomenologic predictions for preonic SUSY without
consideration of specific models or dynamics.

In this paper, we present general remarks on supersymmetric extensions
of composite models. Supersymmetric extensions of fermion-scalar and
three-fermion preonic models are discussed in Sections II.A and II.B,
respectively. Finally we give some concluding remarks in Section III.

\section{Supersymmetric extens\i ons of compos\i te models}

In general, composite models of the SM fermions can be divided into
two different main classes regarding their fundamental constituents:
fermion-scalar and three-fermion models. In this Section we assume
that supersymmetry is realized at preonic level. In this case supersymmetry
can be implanted into models by to ways. The first approach is only
setting the quantum numbers of the preons of an existing model to
make each pair of them superpartners themselves, whereas the second
approach is proposing new superpartners of the preons without any
modifications of existing preons of a model. The first approach is
valid only for fermion-scalar models while the latter can be applied
for three-fermion models also. Below we emphasize some phenomenological
points on supersymmetric extensions of these two type of models.

\subsection{Supersymmetric Fermion-Scalar Models}

Primitive versions of spin-1/2 and spin-0 composite models handled
only SM quarks and specific SM bosons and did not address compositeness
of the SM leptons \cite{key-primitiveFS1,key-primitiveFS2}. However,
as mentioned in the Introduction, the similarity between SM quarks
and leptons leads the idea that if there is a more fundamental level
than SM, then the same particles in that level should give rise to
all SM fermions. Fritzsch and Mandelbaum proposed a composite model
that consists of two fermionic and two scalar color-triplet preons
in 1981 \cite{key-FMmodel} with this point of view. All of the first
generation SM fermions are constructed from minimal invariants $q,\,l=\{FS\}$.
In this case, color singlet SM leptons are predicted to be bound states
of one fermionic preon and one scalar anti-preon

\begin{eqnarray}
l & = & (F\bar{S})=1\oplus8,
\end{eqnarray}

$\vphantom{}$

\noindent with a color-octet partner $l_{8}$. Quarks are expected
to be composed of one fermionic and one scalar anti-preons in a similar
manner 

\begin{eqnarray}
q & = & (\bar{F}\bar{S})=3\oplus\bar{6},
\end{eqnarray}

$\vphantom{}$

\noindent which means that each SM quark has one anti-sextet partner
$\bar{q}_{6}$. 

An extension of the model, with possible electric charge set schemes
under an assumption $|Q_{F,S}|\leq1$, was given by Celikel et al.
and production of the predicted color multiplet particles were studied
from the LHC point of view \cite{key-CelikelKantarSultansoy}. The
same study also includes an option that corresponds to Fritzsch-Mandelbaum
model and one of the options implies the fermion-scalar symmetry from
electric charge viewpoint which may be an indication of super-symmetry
at preonic level. 

If SM fermions are composites of scalar and fermionic preons then
each SM fermion ($FS$) with $m\approx0$ has four partners in a minimal
supersymmetric extension: 

scalar ($\tilde{F}S$) with $m\approx M_{SUSY}$ , 

scalar and vector ($F\tilde{S}$) with $m\approx M_{SUSY}$ , 

and a fermion ($\tilde{F}\tilde{S}$ ) with $m\approx2M_{SUSY}$ ,

\noindent where $M_{SUSY}$ denotes the SUSY scale. 

\subsection{Supersymmetric Three-Fermion Models}

As an example of three-fermion composite models we consider so called
Rishon models \cite{key-Rishon1,key-Rishon2,key-Rishon3}. Harari
and Shupe independently suggested preons as colorless objects first
and suggested that SU(3) color concept arises with combinations of
preons \cite{key-Rishon1,key-Rishon2}. They avoided to explain the
hypercolor dynamics and spin-3/2 SM level fermionic states as mentioned
in their original paper. Latter three-fermion studies came up with
the idea, in which two types of Rishons are assumed to be neutral
and $e/3$ electric charged, color triplet and anti-triplet and SU(3)
hypercolored spin-1/2 preons: V and T, respectively \cite{key-Rishon3}.
According to this class of composite models, SM fermions consist of
three fermionic preons $q,\,l=\{F_{1}F_{2}F_{3}\}$, and therefore,
color singlet SM leptons are predicted to be lighter bound states
of color triplet preons with two color octet and one decuplet partners: 
\begin{center}
\begin{eqnarray}
l=(FFF) & = & 1\oplus8\oplus8\oplus10,
\end{eqnarray}
\par\end{center}

\noindent whereas color triplet SM quarks consist of two preons and
one anti-preon and have color multiplet partners as given below: 
\begin{center}
\begin{eqnarray}
q=(FF\bar{F}) & = & 3\oplus\bar{3}\oplus\bar{6}\oplus15.
\end{eqnarray}
\par\end{center}

If each preon $F_{i}$ has a superpartner $\tilde{F}_{i}$ in a minimal
supersymmetric extension then each composite fermion $(F_{1}F_{2}F_{3})$
with $m\approx0$ should have ten partners: 

three scalars and three vectors ($\tilde{F}_{1}F_{2}F_{3}$), ($F_{1}\tilde{F}_{2}F_{3}$),
($F_{1}F_{2}\tilde{F}_{3}$) with $m\approx M_{SUSY}$,

three fermions ($\tilde{F}_{1}\tilde{F}_{2}F_{3}$), ($\tilde{F}_{1}F_{2}\tilde{F}_{3}$),
($F_{1}\tilde{F}_{2}\tilde{F}_{3}$) with $m\approx2M_{SUSY}$,

and a scalar ($\tilde{F}_{1}\tilde{F}_{2}\tilde{F}_{3}$) with $m\approx3M_{SUSY}$.

\section{Conclusion}

In this study we emphasize some general remarks on supersymmetric
extensions of fermion-scalar and three-fermion composite models since
we believe that supersymmetry should be realized at a more fundamental
level than SM. These extensions predict the existence of new SM fermion-partners
with masses up to three times of SUSY scale. Similar to the previous
composite models, color-multiplet leptons and quarks, which are expected
to have masses much lower than the compositeness scale, are predicted
\cite{key-minimalFS}. Observation of these particles could provide
first indications of supersymmetric preonic models. Of course, mixings
between quarks (leptons, squarks, sleptons) can drastically change
the simple mass relations given in Section II. Therefore, it is quite
possible that the search for SUSY \& compositeness at future colliders
will give rather surprising results. Finally, a more extensive study
on the same topic is in preparation.
\begin{acknowledgments}
Authors are grateful to Ümit Kaya for useful discussion. This study
is supported by TUBITAK under the grant no 114F337.
\end{acknowledgments}

\end{document}